\documentclass[a4paper,11pt]{article}
\usepackage[utf8]{inputenc}
\usepackage[english]{babel}
\usepackage{amsmath, amssymb,graphicx}
\usepackage{braket}
\usepackage{yfonts}
\usepackage{caption}
\usepackage{subcaption}
\usepackage{ulem}
\pdfoutput=1
\usepackage{jheppub}

\title{An Approach to Quantum 2D Gravity}

\author[a,b]{Vladimir V. Belokurov}
\author[a]{and Evgeniy T. Shavgulidze}

\affiliation[a]{Lomonosov Moscow State University,}

\affiliation[b]{Institute for Nuclear
Research of the Russian Academy of Sciences, }

\emailAdd{vvbelokurov@yandex.ru}
\emailAdd{shavgulidze@bk.ru}

\abstract
{ We consider a model of 2D gravity with the action quadratic in curvature and
 represent path integrals  as integrals over the $SL(2,\, \mathbb{R})$ invariant Gaussian functional measure.
We reduce these path integrals to the products of Wiener path integrals and calculate the correlation function of the metric in the first perturbative order.}

\keywords{2D Gravity, Path integrals, Wiener measure}

\arxivnumber{.}

\begin{document}
\maketitle

\section{ Introduction}

\vspace{0.5cm}

The enormous popularity of 2D gravity in the last several decades motivated by its role in string theory and studies of BH physics in
the dimensional reduction approach has grown after realizing the Schwarzian nature of the JT dilaton gravity and the relation of this theory to
SYK model (for some pedagogical reviews see, e.g., \cite{(Ginsparg)}-\cite{(Sarosi)}).

The general form of the 2D gravity action up to the terms quadratic in curvature $K$ is
\begin{equation}
   \label{TotAction}
\tilde{\mathcal{A}}   =
c_{0}\,\int\,\sqrt{\mathcal{G}}\,d^{2}x+c_{1}\,\int\,K\,\sqrt{\mathcal{G}}\,d^{2}x+c_{2}\,\int\,K^{2}\,\sqrt{\mathcal{G}}\,d^{2}x\,.
\end{equation}

The first two terms do not determine the dynamics of 2D gravity.
While the part of the action quadratic in the Gaussian curvature does. It is the model we consider in this paper.

Commonly it is transformed to the dilaton gravity action (see, e.g., the review \cite{(Vass0)}). An alternative way is to deal only with the geometric structures of the surface.
(Some examples of this approach are given in \cite{(KatanVol1)}-\cite{(Vass4)}).

The action (\ref{TotAction}) is invariant under general coordinate transformations. Here, we reduce the set of coordinate transformations and consider the action restricted to the conformal gauge, where the metric of the 2D surface  looks like
\begin{equation}
   \label{metric}
dl^{2} =g(u,\,v)\,\left(du^{2}+dv^{2}\right)=g(z,\,\bar{z})\,dz\,d\bar{z}\,\ \ \ \ \ \,\sqrt{\mathcal{G}}=g\,.
\end{equation}

The Gaussian curvature of the surface is \cite{(Dubrovin)}
\begin{equation}
   \label{curv}
 K=-\frac{1}{2g}\,\Delta\,\log g\,,
\end{equation}
where $\Delta$ stands for the Laplacian.

Note that the action
(\ref{TotAction})
 is invariant under the complex analytic substitutions
 \begin{equation}
   \label{substitution}
z=\chi(\zeta)\,,\ \ \ \ \chi\,g(z,\,\bar{z})=g\left(\chi(\zeta),\,\bar{\chi}(\bar{\zeta}) \right)\,|\chi'_{\zeta}|^{2}\,, \ \ \ \ \ dzd\bar{z}=|\chi'_{\zeta}|^{2}\,d\zeta d\bar{\zeta}\,.
\end{equation}
Therefore, we suppose the integration in the action (\ref{TotAction}) to be over the disc $d\,:\,(|z|\leq 1)\,.$

\vspace{0.5cm}

\section{$SL(2,\, \mathbb{R})$ invariant Gaussian measure }
\label{sec:GaussianMeasure}

\vspace{0.5cm}

We consider the specific form the action (\ref{TotAction})
\begin{equation}
   \label{action}
A= \frac{\lambda^{2}}{2}\,\int\limits_{d}\,\left(K+4 \right)^{2}\,g(z,\,\bar{z})\,dz\,d\bar{z}=\frac{\lambda^{2}}{2}\,\int\limits_{d}\,\left(\Delta \psi\right)^{2}\,\,dz\,d\bar{z}
\end{equation}
where
\begin{equation}
   \label{q}
\Delta \psi=q\,\Delta \log q +\frac{4}{q}\,,\ \ \ \ \ \ \ q=\frac{1}{\sqrt{g}}\,.
\end{equation}

Now path integrals in the theory
\begin{equation}
   \label{PIg}
\int\,\tilde{F}(g)\,\exp\{-\tilde{\mathcal{A}}(g) \}\,dg
\end{equation}
 are path integrals
 \begin{equation}
   \label{PIpsi}
\int\,F(\psi)\,\exp\{-A(\psi) \}\,d\psi
\end{equation}
 over the Gaussian functional measure
\begin{equation}
   \label{measure}
\mu_{\lambda}(d\psi)=\frac{\exp\{-A(\psi) \}\,d\psi}{\int\,\exp\{-A(\psi) \}\,d\psi }\,.
\end{equation}

The extremum of the action (\ref{action}) is given by the equation $\Delta \psi=0\,,$ or
\begin{equation}
   \label{extremum}
 q\,\Delta \log q +\frac{4}{q}=0\,.
\end{equation}

We choose the boundary condition corresponding to the Poincare model of the Lobachevsky plane \cite{(Dubrovin)}
$$
 q_{0}|_{|z|=1}=0\,.
$$
The unique solution in the disk $d\, (|z|\leq 1)$ satisfying the boundary condition is
\begin{equation}
   \label{q0}
 q_{0}=1-z\bar{z}\,.
\end{equation}

Let us rewrite the action (\ref{action}) substituting
$\,\psi\rightarrow f\,$ with
\begin{equation}
   \label{Q}
\Delta \psi= L\,\left[f \right]
\end{equation}
where $L$ is the differential operator
\begin{equation}
   \label{L}
L\equiv q_{0}\,\Delta-\frac{8}{q_{0}} \equiv \frac{1}{q_{0}}\,T^{-1} \,.
\end{equation}
Note that $T^{-1}\equiv\left(q_{0}^{2}\,\Delta-8\right)$ is the Casimir operator of $SL(2,\, \mathbb{R})$ \cite{(Lang)}.

Now the action (\ref{action}) is written as
the integral over the measure $\frac{dz\,d\bar{z}}{ (1-z\bar{z})^{2}}$ invariant under the action   of the group $SL(2,\, \mathbb{R})$ in the disk \begin{equation}
   \label{actioninv}
A= \frac{\lambda^{2}}{2}\,\int\limits_{d}\,\left(T^{-1}\left[f \right] \right)^{2}\,\frac{dz\,d\bar{z}}{ q_{0}^{2} }\,.
\end{equation}

Therefore, we obtain the $SL(2,\, \mathbb{R})$ invariant Gaussian functional measure\footnote{
In what follows, we will not write down the normalization factor explicitly.}
\begin{equation}
   \label{measureinv}
\mu_{\lambda}(df)=\frac{\exp\{-A(f) \}\,df}{\int\,\exp\{-A(f) \}\,df }\,.
\end{equation}

 Note that the $SL(2,\, \mathbb{R})$ invariance of the measure guarantees the invariance for path integrals
$
\int\,F(f)\,\mu_{\lambda}(df)\,.
$

 \vspace{0.5cm}

\section{An algorithm for path integrals calculation}
\label{sec:algorithm}

\vspace{0.5cm}

In this section, we reduce path integrals over the measure (\ref{measureinv}) to the products of Wiener integrals. Here, it is convenient to use polar coordinates.

After the linear substitution (\ref{Q})
$ \Delta \psi(\varrho,\,\varphi)$ is written as
\begin{equation}
   \label{DeltaPsiQ}
\Delta \psi(\varrho,\,\varphi)= L\,\left[f(\varrho,\,\varphi) \right]
= L_{0}\,\left[f(\varrho,\,\varphi)\right] +\frac{\left(1-\varrho^{2}\right)}{\varrho^{2}}\,\frac{\partial^{2}}{\partial\varphi^{2}}\,f(\varrho,\,\varphi)\,.
\end{equation}
where the differential operator $L_{0}$ is
\begin{equation}
   \label{L0}
L_{0}\equiv
\left(1-\varrho^{2}\right)\,\frac{\partial^{2}}{\partial\varrho^{2}}+\frac{\left(1-\varrho^{2}\right)}{\varrho}\,\frac{\partial}{\partial\varrho}-
\frac{8}{\left(1-\varrho^{2}\right)} \,.
\end{equation}

Consider the Fourier series
\begin{equation}
   \label{FourierDeltaPsi}
\Delta \psi(\varrho,\,\varphi)= x_{0}(\varrho)+\sum\limits_{n=1}^{\infty}\left( x_{n}\,\cos n\varphi+y_{n}\,\sin n\varphi \right)
\end{equation}
and
\begin{equation}
   \label{FourierQ}
f(\varrho,\,\varphi)= a_{0}(\varrho)+\sum\limits_{n=1}^{\infty}\left( a_{n}\,\cos n\varphi+b_{n}\,\sin n\varphi \right)\,.
\end{equation}

The coefficients of the first Fourier series can be expressed in terms of the coefficients of the second series
\begin{equation}
   \label{xa}
x_{n}(\varrho)= L_{0}\,\left[a_{n}(\varrho) \right] -n^{2}\,\frac{\left(1-\varrho^{2}\right)}{\varrho^{2}}\,a_{n}(\varrho)
\end{equation}
and
\begin{equation}
   \label{yb}
y_{n}(\varrho)= L_{0}\,\left[b_{n}(\varrho) \right] -n^{2}\,\frac{\left(1-\varrho^{2}\right)}{\varrho^{2}}\,b_{n}(\varrho) \,.
\end{equation}

Now the action (\ref{action}) is written as
\begin{equation}
   \label{xyaction}
A=\frac{\lambda^{2}}{2}\,2\pi\int\limits_{0}^{1}\,\left( x_{0}(\varrho)\right)^{2}\,\varrho\,d\varrho +\frac{\lambda^{2}}{2}\,\pi\,\sum\limits_{n=1}^{\infty}\,\left[\int\limits_{0}^{1}\,\left( x_{n}(\varrho)\right)^{2}\,\varrho\,d\varrho + \int\limits_{0}^{1}\,\left( y_{n}(\varrho)\right)^{2}\,\varrho\,d\varrho \right]\,.
\end{equation}

After the substitution
\begin{equation}
   \label{aVn}
a_{n}(\varrho)=\varrho^{n}\,\frac{\left(1+n+(1-n)\varrho^{2}\right)}{\left(1-\varrho^{2}\right)}\,\int\limits_{\varrho}^{1}\,
\frac{\left(1-\varrho_{1}^{2}\right)^{2}\,V_{n}(\varrho_{1})}{\varrho_{1}^{2n+1}\,\left(1+n+(1-n)\varrho_{1}^{2}\right)^{2}}\,d\varrho_{1} \,,
\end{equation}
we obtain
\begin{equation}
   \label{ntermaction}
\int\limits_{0}^{1}\,\left( x_{n}(\varrho)\right)^{2}\,\varrho\,d\varrho =\int\limits_{0}^{1}\,\frac{\left(1-\varrho^{2}\right)^{4}}{\varrho^{2n+1}\,\left(1+n+(1-n)\varrho^{2}\right)^{2}}\,\left( V'_{n}(\varrho)\right)^{2}\,d\varrho\,,\ \ \ \ \ n=0,1,\ldots,
\end{equation}
where
\begin{equation}
   \label{Vn}
V_{n}(\varrho)=\varrho^{2n+1}\,\frac{\left(1+n+(1-n)\varrho^{2}\right)^{2}}{\left(1-\varrho^{2}\right)^{2}}
\,\left(\frac{\left(1-\varrho^{2}\right)\,a_{n}(\varrho)}{\varrho^{n}\,\left(1+n+(1-n)\varrho^{2}\right)} \right)'\,.
\end{equation}

Therefore we have
\begin{equation}
   \label{Untermaction}
\int\limits_{0}^{1}\,\left( x_{n}(\varrho)\right)^{2}\,\varrho\,d\varrho =\int\limits_{0}^{+\infty}\,
\left(U'_{n}(\tau_{n})\right)^{2}\,d\tau_{n}
\end{equation}
with
\begin{equation}
   \label{Un}
\tau_{n}=\int\limits_{0}^{\varrho}\,\frac{\varrho_{1}^{2n+1}}{\left(1-\varrho_{1}^{2}\right)^{2}}\,
\left(\frac{2}{\left(1-\varrho_{1}^{2}\right)}+n-1\right)^{2}\,d\varrho_{1}\,,\ \ \ \ \ \ \ \ \ \ U_{n}(\tau_{n})=V_{n}(\varrho)\,.
\end{equation}

The same equations are valid for the other terms in (\ref{xyaction})
\begin{equation}
   \label{tildeUntermaction}
\int\limits_{0}^{1}\,\left(y_{n}(\varrho)\right)^{2}\,\varrho\,d\varrho =\int\limits_{0}^{+\infty}\,
\left(\tilde{U}'_{n}(\tau_{n})\right)^{2}\,d\tau_{n}\,.
\end{equation}

Now the measure $\mu_{\lambda}(df)$ is represented as the product of the Wiener measures\footnote{See the footnote 1 in section (\ref{sec:GaussianMeasure}).}
\begin{equation}
   \label{MeasureProduct}
\mu_{\lambda}(df)=w_{\frac{1}{\lambda\sqrt{2\pi}}}\left(dU_{0}\right)\,\prod\limits_{n=1}^{\infty}\,w_{\frac{1}{\lambda\sqrt{\pi}}}\left(dU_{n}\right)\,
w_{\frac{1}{\lambda\sqrt{\pi}}}\left(d\tilde{U}_{n}\right)
\end{equation}
where
\begin{equation}
   \label{WienerMeasure}
w_{\sigma}\left(dU\right)=\exp\left\{-\frac{1}{2\sigma^{2}}\,\int\limits_{0}^{+\infty}\,\left(U'(\tau)\right)^{2}\,d\tau \right\}\,dU\,.
\end{equation}

For path integrals with integrands that depend on the modes with definite numbers, the product of the measures is reduced because of the cancelation of the same terms in the nominator and in the denominator.

The simplest basic path integrals over the measure $\mu_{\lambda}(df)$ are
\begin{equation}
   \label{VVn}
\int\limits_{C([0,\,1])}\,V_{n}(\varrho_{1})\,V_{n}(\varrho_{2})\,\mu_{\lambda}(df)
\end{equation}
with
$$
0\leq \varrho_{1}\leq \varrho_{2}<1\,,\ \ \ \ \ \ \ \ \ \ \tau_{n\,(1)}= \tau_{n}(\varrho_{1})\leq \tau_{n}(\varrho_{2})=\tau_{n\,(2)}\,.
$$
For $n=0$, (\ref{VVn}) is equal to\footnote{See, e.g., \cite{(Kuo)}})
$$
\int\limits_{C([0,\,1])}\,V_{0}(\varrho_{1})\,V_{0}(\varrho_{2})\,\mu_{\lambda}(df)=\int\limits_{C([0,\,+\infty)}\,U_{0}(\tau_{0\,(1)})\,U_{0}
(\tau_{0\,(2)})\,w_{\frac{1}{\lambda\sqrt{2\pi}}}\left(dU_{0}\right)
$$
$$
=\frac{1}{2\pi\lambda^{2}}\,\tau_{0\,(1)}
$$
\begin{equation}
   \label{UU0}
=
\frac{1}{2\pi\lambda^{2}}\,\int\limits_{0}^{\varrho_{1}}\frac{\varrho}{(1-\varrho^{2})^{2}}\left(\frac{2}{1-\varrho^{2}}-1 \right)^{2}d\varrho=
\frac{1}{24\pi\lambda^{2}}\,\left(\frac{2}{1-\varrho_{1}^{2}}-1 \right)^{3}\,.
\end{equation}
For $n\geq 1$,
$$
\int\limits_{C([0,\,1])}\,V_{n}(\varrho_{1})\,V_{n}(\varrho_{2})\,\mu_{\lambda}(df)
=\int\limits_{C([0,\,+\infty)}\,U_{n}(\tau_{n\,(1)})\,U_{0}(\tau_{n\,(2)})\,w_{\frac{1}{\lambda\sqrt{\pi}}}\left(dU_{n}\right)
$$
\begin{equation}
   \label{UUn}
=\frac{1}{\pi\lambda^{2}}\,\tau_{n\,(1)}
=
\frac{1}{\pi\lambda^{2}}\,\int\limits_{0}^{\varrho_{}}\frac{\varrho^{2n+1}}{(1-\varrho^{2})^{2}}\left(\frac{2}{1-\varrho^{2}}+n-1 \right)^{2}d\varrho\,.
\end{equation}

More complicated path integrals with higher powers of integration variables are calculated using the Wick theorem.

\vspace{0.5cm}

\section{Correlation function of the metric}
\label{sec:algorithm}

\vspace{0.5cm}

In this section, we calculate the correlation function of the metric given by the integral
\begin{equation}
   \label{corrfunct}
<\, g(\varrho_{1},\,\varphi_{1})\, g(\varrho_{2},\,\varphi_{2})\,>_{\mu}=\int\limits_{C(d)}\, g(\varrho_{1},\,\varphi_{1})\, g(\varrho_{2},\,\varphi_{2})\ \mu_{\lambda}(df)
\end{equation}
in the first nontrivial  perturbative order.

Due to the $SL(2,\, \mathbb{R})$ invariance,
the correlation function of the metric can be rewritten as
$$
<\, g(\varrho_{1},\,\varphi_{1})\, g(\varrho_{2},\,\varphi_{2})\,>_{\mu}
=\frac{q^{2}_{0}(\varrho_{\ast})}{q^{2}_{0}(\varrho_{1})\,q^{2}_{0}(\varrho_{2})\,}\,\int\limits_{C(d)}\, g(0,\,0)\, g(\varrho_{\ast},\,\varphi_{\ast})\ \mu_{\lambda}(df)
$$
\begin{equation}
   \label{corrfunctast}
=\frac{q^{2}_{0}(\varrho_{\ast})}{q^{2}_{0}(\varrho_{1})\,q^{2}_{0}(\varrho_{2})\,}\,<\, g(0,\,0)\, g(\varrho_{\ast},\,\varphi_{\ast})\,>_{\mu}\,,
\end{equation}
where $(0,\,0)$ and $(\varrho_{\ast},\,\varphi_{\ast}) $ are the results of the shift at the Lobachevsky plane of the coordinates
$(\varrho_{1},\,\varphi_{1}) $ and $(\varrho_{2},\,\varphi_{2})\ \ (0\leq\varrho_{\ast}\leq 1\,,\ \ 0\leq\varphi_{\ast}\leq 2\pi)\,. $
In particular,
\begin{equation}
   \label{phoast}
\varrho_{\ast}=\frac{\sqrt{\varrho_{1}^{2}+\varrho_{2}^{2}-2\varrho_{1}\varrho_{2}\cos\left( \varphi_{2}-\varphi_{1}\right)}}{\sqrt{1+\varrho_{1}^{2}\varrho_{2}^{2}-4\varrho_{1}\varrho_{2}\cos\left( \varphi_{2}-\varphi_{1}\right)}}\,.
\end{equation}

The integration measure is invariant under rotations and averaged values depend only on the variable $\varrho\,.$

Let us represent the function $q(\varrho,\,\varphi)$ as the series in the powers of the variable $f$
\begin{equation}
   \label{qseries}
q(\varrho,\,\varphi)= q_{0}(\varrho)\,\left(1+p_{1}\left(f(\varrho,\,\varphi)\right)+p_{2}\left(f(\varrho,\,\varphi)\right)+\ldots\right)\,,
\end{equation}

From (\ref{q}) and (\ref{Q}), it follows that\footnote{Only the terms essential in the studied perturbative order $\left(\sim \frac{1}{\lambda^{2}} \right)$ are explicitly written down.}
\begin{equation}
   \label{p1}
p_{1}\left(f\right)=f
\end{equation}
and
\begin{equation}
   \label{p2}
L\,\left[p_{2} -\frac{1}{2}f^{2}\right]=f\,L\,\left[f\right]-\frac{8}
{q_{0}}\,f^{2}\,.
\end{equation}

For first nontrivial order calculations, we expand the metric up to the quadratic in $f$ terms
\begin{equation}
   \label{gsecondorder}
g(\varrho_{\ast},\,\varphi_{\ast})= \frac{1}{q^{2}_{0}(\varrho_{\ast})}\left(1-2 f(\varrho_{\ast},\,\varphi_{\ast})
+3f^{2}(\varrho_{\ast},\,\varphi_{\ast})
-2\frac{p_{2}\left(f(\varrho_{\ast},\,\varphi_{\ast})\right)}{q_{0}(\varrho_{\ast})}\right)
\end{equation}
where $p_{2}$ is given by (\ref{p2}) and  $(\varrho_{\ast},\,\varphi_{\ast}) $ is a point inside the disk $d$ .

Note that, due to the $SL(2,\, \mathbb{R})$ invariance,
\begin{equation}
   \label{averagedmetric}
<\, g(\varrho,\,\varphi)\,>_{\mu}=\int\limits_{C(d)}\, g(\varrho,\,\varphi)\ \mu_{\lambda}(df)
=\frac{1}{q^{2}_{0}(\varrho)}\,<\, g(0,\,0)\,>_{\mu}\,.
\end{equation}

Taking into account (\ref{averagedmetric}) and (\ref{gsecondorder}), we obtain
\begin{equation}
   \label{ggff}
<\, g(0,\,0)\, g(\varrho_{\ast},\,\varphi_{\ast}) \,>_{\mu}
=\frac{2}{q^{2}_{0}(\varrho_{\ast})}\,\left\{<\, g(0,\,0)\,>_{\mu}+2<\, f(0,\,0)\, f(\varrho_{\ast},\,\varphi_{\ast}) \,>_{\mu}\right\}\,.
\end{equation}

Now we calculate the first term in the sum in the r.h.s. of  (\ref{ggff})
\begin{equation}
   \label{g0}
<\, g(0,\,0)\, >_{\mu}\equiv <\, g\, >_{\mu}(0)=1+3<\, f^{2}\, >_{\mu}(0)-2<\,p_{2}\, >_{\mu}(0)
\,.
\end{equation}

Using the $SL(2,\, \mathbb{R})$ invariance  of the integration measure, we  obtain from  (\ref{p2})
\begin{equation}
   \label{p2avL0}
<\,p_{2}\,>_{\mu}(0)=\frac{1}{2}<\,f^{2}\,>_{\mu}(0)-8<\,f^{2}\,>_{\mu}(0)\
L_{0}^{-1}\,\left[ \frac{1}{q_{0}}\right](0)\,,
\end{equation}
with
$$
L_{0}^{-1}\left[h(\varrho_{1}) \right](\varrho)=-(1+\varrho^{2})\,
\int\limits_{\varrho}^{1}\,\frac{(1-\varrho_{2}^{2})^{2}}{\varrho_{2}\,(1+\varrho_{2}^{2})^{2}} \,\int\limits_{0}^{\varrho_{2}}\,\frac{\varrho_{1}\,(1+\varrho_{1}^{2})}{(1-\varrho_{1}^{2})^{2}}\,h(\varrho_{1}) \,d\varrho_{1}\,d\varrho_{2}\,.
$$
Obviously,
$$
L_{0}^{-1}\,\left[ \frac{1}{q_{0}}\right](0)=-\frac{1}{8}\,,
$$
and therefore
\begin{equation}
   \label{g0}
 <\, g\, >_{\mu}(0)=1\,.
\end{equation}

The second term in (\ref{ggff}) is reduced to
\begin{equation}
   \label{ffaa}
<\, f(0,\,0)\, f(\varrho_{\ast},\,\varphi_{\ast}) \,>_{\mu}
=<\, a_{0}(0)\, f(\varrho_{\ast},\,\varphi_{\ast}) \,>_{\mu}=<\, a_{0}(0)\, a_{0}(\varrho_{\ast}) \,>_{\mu}
\end{equation}
and can be calculated using the technique of the previous section.

Thus, the path integration in (\ref{ffaa}) results in
$$
<\, f(0,\,0)\, f(\varrho_{\ast},\,\varphi_{\ast}) \,>_{\mu}
$$
$$
=\frac{1}{6\pi\lambda^{2}}\,\int\limits_{\varrho}^{1}\,\frac{(1-\varrho_{2}^{2})^{2}}{\varrho_{2}\,(1+\varrho_{2}^{2})^{2}} \,\int\limits_{\varrho_{\ast}}^{\varrho_{2}}\,\frac{\varrho_{1}\,(1+\varrho_{1}^{2})}{(1-\varrho_{1}^{2})^{2}}\,\frac{\varrho_{1}^{2}
\,(3+\varrho_{1}^{4})}
{(1-\varrho_{1}^{2})^{3} } \,d\varrho_{1}\,d\varrho_{2}
$$
$$
+\frac{1}{12\pi\lambda^{2}}\,\int\limits_{\varrho}^{1}\,\frac{(1-\varrho_{2}^{2})^{2}}{\varrho_{2}\,(1+\varrho_{2}^{2})^{2}} \,d\varrho_{2} \,\int\limits_{0}^{\varrho_{\ast}}\,\frac{\varrho_{1}\,(1+\varrho_{1}^{2})}{(1-\varrho_{1}^{2})^{2}}\,\frac{\varrho_{1}^{2}\,(3+\varrho_{1}^{4})}
{(1-\varrho_{1}^{2})^{3} } \,d\varrho_{1}
$$
$$
=\frac{1}{12\pi\lambda^{2}}\,\left\{ -\frac{1}{2}\int\limits_{\varrho_{\ast}}^{1}\,\frac{\log t}{1-t}dt+2\log2+\frac{1}{2}
-\frac{1}{1+\varrho_{\ast}^{2}}+\frac{1}{1+\varrho_{\ast}^{2}}\log\left(\frac{1-\varrho_{\ast}^{2}}{\varrho_{\ast}} \right)\right.
$$
\begin{equation}
   \label{result}
  \left. +\log\varrho_{\ast}-2\log(1+\varrho_{\ast}^{2})-\log\sqrt{1-\varrho_{\ast}^{2}}+\log\varrho_{\ast}\,\log\sqrt{1-\varrho_{\ast}^{2}}\
 \right\}\,.
\end{equation}

In the limit $\varrho_{\ast}\rightarrow 1\,,$ we have an interesting (and expected) physical result. The quantum corrections to the correlation function of the metric tend to zero and\footnote{The factor $2$ stands because of the symmetry $\varrho_{\ast}\Leftrightarrow 0$.}
$$
<\, g(0,\,0)\, g(\varrho_{\ast},\,\varphi_{\ast}) \,>_{\mu}\ \ \rightarrow \ \  <\, g(0,\,0)\,>_{\mu}\,<\, g(\varrho_{\ast},\,\varphi_{\ast}) \,>_{\mu}\,.
$$

\vspace{0.5cm}

\section{Conclusion}
\label{sec:concl}

\vspace{0.5cm}

In this paper, we consider 2d gravity with the action quadratic in the Gaussian curvature of the surface.
We rewrite the theory in the form
invariant under the action   of the group $SL(2,\, \mathbb{R})$ in the disk and then we
treat path integrals in the theory as integrals over the $SL(2,\, \mathbb{R})$ invariant Gaussian functional measure.
Then we demonstrate how these path integrals can be reduced to the products of Wiener path integrals and, as an example, calculate the correlation
function of the metric in the first perturbation theory order.

In \cite{(BShPIQG)}, we reduced path integrals of quadratic gravity in FLRW metric to Wiener path integrals.
The elaborated in this paper method may be considered as a first step in generalizing the approach of \cite{(BShPIQG)} to higher dimensions.

In some other context, path integrals related with  Casimir operator of $SL(2,\, \mathbb{R})$ where studied in \cite{(SL2Rorbits)}.
However the measure used in \cite{(SL2Rorbits)} $\exp\left\{\left(T^{-1}h,\,h \right) \right\}dh $ differs from the measure obtained in our paper
$\exp\left\{\left(T^{-1}h,\,T^{-1}h \right) \right\}dh $ (see (\ref{measureinv}), (\ref{actioninv})).

\end{document}